\newcommand{\beq}{\begin{equation}}
\newcommand{\eeq}{\end{equation}}
\newcommand{\bed}{\begin{displaymath}}
\newcommand{\eed}{\end{displaymath}}
\newcommand{\beqa}{\begin{eqnarray}}
\newcommand{\eeqa}{\end{eqnarray}}
\newcommand{\beqan}{\begin{eqnarray*}}
\newcommand{\eeqan}{\end{eqnarray*}}

\newcommand{\goto}{\rightarrow}

\newcommand{\fn}{\footnote}
\newcommand{\bfl}{\begin{flushleft}}
\newcommand{\efl}{\end{flushleft}}
\newcommand{\bfr}{\begin{flushright}}
\newcommand{\efr}{\end{flushright}}

\newcommand{\lae}{\mathrel{<\kern-1.0em\lower0.9ex\hbox{$\sim$}}}
\newcommand{\gae}{\mathrel{>\kern-1.0em\lower0.9ex\hbox{$\sim$}}}
\newcommand{\noi}{\noindent}
\newcommand{\btab}{\begin{tabbing}}
\newcommand{\etab}{\end{tabbing}}

\newcommand{\mn}{{\rm min}}

\newcommand{\non}{\nonumber}
\newcommand{\lb}{\langle}
\newcommand{\rb}{\rangle}
\newcommand{\bfig}{\begin{figure}}
\newcommand{\efig}{\end{figure}}

\documentstyle[emulateapj,apjfonts,psfig]{article}
\slugcomment{{\it Submitted to ApJ 03/15/02; accepted 05/15/02}}

\begin{document}
\title{SHAPES OF MOLECULAR CLOUD CORES AND THE FILAMENTARY 
MODE OF STAR FORMATION}
\author{CHARLES L.\ CURRY\altaffilmark{1}}
\affil{Department of Physics, 
University of Waterloo, Waterloo, ON  N2L 3G1}
\affil{{\rm and}} 
\affil{Department of Physics and Astronomy, 
University of Western Ontario, London, ON  N6A 3K7}
\altaffiltext{1}{Email: curry@astro.uwaterloo.ca} 

\begin{abstract} 
Using recent dust continuum data, we generate the intrinsic 
ellipticity distribution of dense, starless molecular cloud cores. 
Under the hypothesis that the cores are all either oblate or 
prolate randomly-oriented spheroids, we show that a satisfactory 
fit to observations can be obtained with a gaussian prolate 
distribution having a mean intrinsic axis ratio of 0.54. 
Further, we show that correlations exist between the apparent 
axis ratio and both the peak intensity and total flux density 
of emission from the cores, the sign of which again favours 
the prolate hypothesis. The latter result shows that the mass 
of a given core depends on its intrinsic ellipticity. 
Monte Carlo simulations are performed to find the 
best-fit power law of this dependence. Finally, we show 
how these results are consistent with an evolutionary scenario leading 
from filamentary parent clouds to increasingly massive, condensed,  
and roughly spherical embedded cores. 
\end{abstract}
\keywords{ISM: clouds --- ISM: structure --- Stars: formation}

\section{INTRODUCTION}
\label{sec-intro}
Little is known with certainty about the intrinsic shapes of the dense 
molecular clouds that give birth to stars. The present situation is 
reminiscent of the analogous study of elliptical galaxies some thirty 
years ago, when relatively 
few tests had been performed on limited datasets. In that field, 
attempts to model the observed distributions of ellipticity, surface 
brightness, and velocity dispersion with axisymmetric objects 
(e.g., oblate or prolate spheroids) met with limited success 
(Merritt 1982). Additional kinematical data and $N$--body 
simulations have led to the conclusion that many ellipticals 
are, in fact, triaxial (Merrifield and Binney 1998). 

While from a modelling perspective our understanding of molecular 
cloud shapes lags behind that of stellar systems, much can be 
learned by applying the same methods in this new arena. At the same 
time, it is important to keep in mind certain salient differences 
between the two contexts, aside from the obvious difference of
scale. First and foremost, gaseous 
self-gravitating clouds are not collisionless systems. Agents such as 
thermal instability, pressure gradients, and magnetic fields, 
more or less unique to the ISM context, have all been shown 
to be important in various physical regimes. Second, 
studies of the most centrally condensed cores of molecular clouds 
reveal them to be not the isolated (or infinite) balls of dense gas
considered by simple theory, but rather the lowest rung in a hierarchy 
of structure beginning with parsec-scale entities (giant molecular 
clouds). This aspect needs to be considered when theory is brought 
to bear upon intrinsic core properties. 

Clues to core structure are beginning to be extracted from
morphological studies. Early analyses were hampered by 
datasets of limited size, which reduced the statistical significance 
of the conclusions (David \& Verschueren 1987; Myers et al.\ 1991; Ryden 
1996). Nevertheless, under the hypothesis that each object is a spheroid 
randomly-oriented to the line of sight, each of these studies concluded 
that cores were more likely to be intrinsically prolate than oblate. 
The recent study of Jones, Basu, \& Dubinski (2001) analyzed the
largest dataset to date: 264 ammonia cores compiled by 
Jijina, Myers, \& Adams (1999). 
By showing that the best-fit probability distributions 
of prolate and oblate spheroids became negative near $p=1$, these authors 
rejected the hypothesis of axisymmetry altogether. A closer examination 
of the analysis technique, however, gives a likely explanation of their 
results (see below \S \ref{sec-Nvsp}).

All of the aforementioned analyses utilized but one observational 
diagnostic of core shape: the projected axis ratio, or ellipticity.
However, as remarked by Fleck (1992), the {\it mean} value of this 
quantity cannot be used to distinguish one spheroidal shape over 
another. Moreover, 
as pointed out by Binney \& de Vaucouleurs (1981) and Ryden (1996), the 
measurement of axis ratios is subject to various systematic biases that  
can affect the overall distribution and subsequent analysis. In fact, 
other physical properties---such as column density, velocity dispersion, 
and mass---have influenced theoretical models of cloud cores far more than 
have the perceived ellipticities. We feel that to abandon the 
hypothesis of axisymmetry in lieu of examining these diagnostics is 
premature, especially given the lack of a physical basis for 
triaxiality in the ISM context (akin to the anisotropic velocity 
distribution of stars in elliptical galaxies). 

In \S \ref{sec-observ} and \ref{sec-intrins}, we employ a number of tests 
to discriminate between the oblate and prolate spheroidal hypotheses. 
Two of the 
tests are new in the ISM context, and the other---the distribution of
projected axis ratios---has not previously been applied to the dust 
continuum data on which we base our analysis (however, subsets of 
these data were recently analyzed by Jones \& Basu (2002), under the 
oblate and triaxial hypotheses only). Further, by simulating the
observed sample using distributions of model cores of both types, we 
are able to place rather stringent constraints on one intrinsic
property, the variation of polar intensity with intrinsic
ellipticity. All of the tests are 
independent of distance, still a very uncertain quantity for these 
objects (see, e.g., Launhardt \& Henning 1997).

Finally, in \S \ref{sec-discuss} we interpret our results in the 
framework of an evolutionary 
sequence of core shapes as a function of time. In this picture, core 
morphology at early times is primarily determined by the nature 
of the surrounding nonisotropic mass distribution. As the core grows 
in mass, it approaches a spherical shape consistent with the dominance 
of self-gravity. As we shall see, this behavior is consistent with the 
present observational picture. 

\section{DATA}
\label{sec-data}
We analyzed recent dust continuum (submillimeter) observations of dense 
cores in three prominent star-forming regions: Orion OMC-2 and 3 (Chini et 
al.\ 1997), $\rho$ Ophiuchus (Motte, Andr\'e, \& Neri 1998), and Orion 
B (Motte et al.\ 2001). Collectively, we refer to these observations, 
which identified a total of 165 dense cores, 
as the ``combined continuum dataset.'' The dust continuum data have at 
least two distinct advantages over molecular line measurements as a 
probe of core structure. First, while molecular lines have a limited 
range of both column and number density sensitivity (a factor $\sim$ 
10--30) due to opacity and molecular freeze-out onto grains, thermal 
emission from dust is presumably optically thin. This means that column 
(or continuum flux) densities are bounded only from below by detector 
sensitivity, while 
the derived number densities vary inversely with beam resolution 
(beam-averaged H$_2$ column densities $\gae 10^{23}$ cm$^{-2}$ are 
typical; e.g., Testi \& Sargent 1998).  
Second, while there is in most cases a rough 
spatial coincidence between the peaks seen in the dust continuum  
surveys and those seen in dense molecular tracers, the continuum 
cores may in fact be more representative of the progenitors 
of stars. This is suggested by the remarkable agreement between the 
mass spectrum of the cores and the field star initial mass function 
(Testi \& Sargent 1998; Motte et al.\ 1998, 2001; Johnstone et al.\ 
2000, 2001).

Since we make extensive use of the Motte et al.\ (2001) dataset 
(hereafter M01), we note here some of its properties used later in 
the analysis. The survey covered a $32\arcmin \times 18\arcmin$ region 
of Orion B, including the NGC 2068/2071 protoclusters, and identified 82 
dense condensations. We analyzed the $850 \micron$ data, 
for which the half-power beam width is $R_B \simeq 13\arcsec
\approx 5000$ AU at the assumed distance of Orion B (400 pc). The 
one-sigma rms noise within the beam is $\Delta I = 22$ mJy/beam. For 
a core with FWHM size $R$, the rms noise in the integrated (total) 
flux density $S$ is $\Delta S \approx 22$ mJy $\times \sqrt{(R^2 + 
R_B^2)/R_B^2} 
\times 2 = 65$ mJy for $R \simeq 5500$ AU, the mean core size in the 
sample. The absolute calibration uncertainty of the measurements is 
$\sim$ 20\%. Only the total flux for each core was 
given in M01; peak intensities $I$ (flux within a single beam) were 
obtained from the authors. The total error in log $I$ 
ranges between $-0.22$ and $+0.14$ over the range observed. We take 
the error in log $p$ to be $\pm 0.05$, as determined from the precision 
of M01's figures for the core major and minor axes ($\pm 200$ AU).

While several dust continuum surveys of other regions are now available, 
we have used only those for which both apparent major/minor axes and 
peak intensities were tabulated. The reason for this will become 
clear presently.     

\section{THE DISTRIBUTION OF CORE ELLIPTICITIES}
\label{sec-Nvsp}
The frequency distribution of the apparent axis ratio $p$ 
(apparent minor/major axis length $\equiv b'/a'$) is the only tool 
used thus far in efforts to deduce the intrinsic shapes of molecular 
cloud cores. While prone to selection effects at both low and high 
ellipticity, the method can give some insight into the constraints 
involved (see \S \ref{sec-intro} for references). The histogram in 
Figure 1$a$ shows the distribution of 
core apparent axis ratios, using the combined continuum dataset.
The mean value is $\lb p \rb = 0.63$ (median: 0.64),  
with a standard deviation of 0.20. 
This distribution is similar to that seen in the extensive 
and much more heterogeneous ammonia dataset of Jijina et al.\ 
(1999) (see Jones et al.\ 2001). In particular, both distributions 
are skewed towards $p=1$: in Figure 1 there are more than twice as 
many objects 
with $0.8 \leq p \leq 1$ than with $0.1 \leq p \leq 0.3$. This 
characteristic is particularly evident in the Chini et al.\ and Motte 
et al.\ (1998) samples, in which 86\% and   
79\% of the cores, respectively, have $p \geq 0.5$. 

Let the intrinsic frequency distribution of cores of intrinsic axis 
ratio $q$ (intrinsic minor/major axis length $\equiv b/a$) be denoted 
by $\psi (q)$. Then, if the cores are oriented randomly on the sky, 
the observed distribution $\phi$ of apparent (projected) axis ratios 
$p$ is given by (Noerdlinger 1979; Fall \& Frenk 1983):
\beqa
\phi (p) &=& p \int_0^p dq (1-q^2)^{-1/2} (p^2 - q^2)^{-1/2} ~\psi (q)
\non \\  
&&~~~~~~~~~~~~~~~~~~~~~~~~~~~~~~~~~~({\rm oblate}); \label{eq-obphi} \\  
&=& p^{-2} \int_0^p dq q^2 (1-q^2)^{-1/2} (p^2 - q^2)^{-1/2} ~\psi (q)
\non \\  
&&~~~~~~~~~~~~~~~~~~~~~~~~~~~~~~~~~~({\rm prolate}).
\label{eq-plphi} 
\eeqa

The derivation of $\psi (q)$ from the observed $\phi (p)$ has been 
attempted in a number of ways. For example, Ryden (1996) employed a 
nonparametric kernel estimator, while Jones et al.\ (2001) used an analytic
inversion method. Both replaced the observed distribution histogram 
with a smooth function before inverting to find $\psi (q)$. 
In the process, restrictions were placed on either the form of $\phi (p)$
(an odd polynomial of degree 5; Jones et al.) or its slope 
at the endpoints ($d\phi/dp = 0$; Ryden). Despite the fact that this 
biases any comparison of the observed and theoretical distributions,
the goodness-of-fit criteria in both studies rely upon the behavior 
near the endpoints, particularly at $p=1$. Thus, these authors' finding 
that the probability distributions of spheroids become negative (for 
some samples) near $p=1$ needs to be re-established in the absence of 
such restrictions. 

While a direct inversion method, such as the type devised by 
Lucy (1974) (and applied in the galactic context by Noerdlinger), 
would be the most robust means of deriving $\psi (q)$, our objective 
here is more modest. We seek merely to demonstrate that a reasonable 
distribution of randomly-oriented spheroids can reproduce the
histogram of Figure 1$a$. We let $\psi (q)$ equal a gaussian with mean 
value $\lb q \rb$ and dispersion $\sigma$, and generated $\phi (p)$ from 
equations (\ref{eq-obphi}) and (\ref{eq-plphi}) for a range of values of 
these parameters. 

The resulting $\chi^2$ minimization fits to the observed histogram (now 
shown as a probability density) are shown in Figure 1$b$. While neither 
prolate nor oblate distributions give a completely satisfactory fit, 
the prolate distribution with $\lb q \rb = 0.54$ and $\sigma = 0.19$ 
($\chi^2 = 0.28$) is clearly superior. 
The fit in the oblate case is less satisfactory: $\lb q \rb = 
0.39$ and $\sigma = 0.16$ with $\chi^2 = 0.51$. In this case, while 
the observed 
$\phi$ can be fit at low $p$, the peak in the derived $\phi$ lies at a
smaller $p$ than observed, and objects with nearly circular 
projections are over-produced. The best-fitting oblate distribution 
consists of cores that are intrinsically more flattened than in the 
prolate case, which also may be considered a priori less likely (Myers 
et al., 1991). While the poor agreement may simply be a result of 
restricting $\psi$ to be gaussian, a less generic form of $\psi$ 
used in the oblate, but not in the prolate, case has little  
justification. 

We also applied the $\chi^2$ analysis to the axis ratio 
distribution of Jijina et al.\ (1999). The same qualitative result 
was found, but with less of a difference between the oblate 
($\chi^2 = 0.97$) and prolate ($\chi^2 = 0.88$) scenarios.\fn{The 
elevated $\chi^2$ in both cases is due to a poor fit near $p=1$, where   
the number of nearly circular objects more than doubles in the 
rightmost bin ($0.9 \leq p \leq 1$). It is likely that the latter
number has been overestimated in this more heterogeneous sample, due 
to rounding errors in older measurements (see, e.g., Ryden 1996).}
These results make clear that, despite recent claims to the contrary, 
the observed $\phi (p)$ is reasonably well-fit by a gaussian
distribution $\psi (q)$ of randomly-oriented prolate spheroids.
This consequently weakens the motivation for triaxial models which, 
in any case, have no theoretical basis in the context of gaseous, 
equilibrium clouds. We shall use these best-fit intrinsic 
distributions to generate peak intensities for simulated 
datasets in \S \ref{sec-intrins}. 

\section{OBSERVED CORRELATIONS}
\label{sec-observ}
\subsection{Relation Between Peak Intensity and Apparent Ellipticity}
\label{sec-Svsp}
Consider a cloud core in which the mass distribution is constant on 
similar ellipsoids, either oblate or prolate. In an optically thin 
tracer, the particular line of sight corresponding to the peak
intensity $I$ (or peak flux density; units Jy/beam) represents the 
complete column through the core.\fn{All intensities and flux densities 
referred to in this work are monochromatic, measured at either 850 
or 1300 $\micron$.} It should therefore depend upon the 
apparent axis ratio $p$. Thus we have (Richstone 1979; Merritt 1982):
\beqa
I(\theta) & = & \frac{q}{p} I_e = p^{-1} I_p ~~~~~~~~~~~~~ ({\rm oblate})
\label{eq-Sob} \\
I(\theta) & = & \frac{p}{q} I_e = p I_p ~~~~~~~~~~~~~~~~ ({\rm prolate}), 
\label{eq-Spl}
\eeqa
where $\theta$ is the angle between the observer's line of sight and 
the equatorial plane, $I_e = I (0)$ is the intensity as seen toward the
same plane, and $I_p = I(\pi/2)$ is the intensity as seen down the polar 
axis. The intrinsic and apparent axis ratios, $q$ and $p$, are related 
to $\theta$ via
\beqa
{\rm cos}^2 \theta &=& \frac{1 - p^2}{1 - q^2} ~~~~~~~~~~~~~~~~~ 
({\rm oblate}) \label{eq-obltheta} \\
{\rm sin}^2 \theta &=& \frac{1 - q^2/p^2}{1 - q^2} ~~~~~~~~~~~ 
({\rm prolate}). 
\label{eq-protheta} 
\eeqa
In both cases, the value $p = 1$ refers to the case in which the object 
is viewed along the polar axis, and therefore has a circular
projection. 

Let us assume, in the first approximation, that: (i) the cores are 
randomly oriented, and (ii) $I_p$ does not vary from one core to 
the next in a population of either all-oblate or all-prolate spheroids. 
Then the observed intensity $I$ will increase towards $p=1$ in the prolate 
case, and decrease towards $p=1$ in the oblate case. Specifically, the 
expected slopes in a (log $I$, log $p$) plot are $+1$ and $-1$, 
respectively (throughout this paper, log denotes log$_{10}$). [For  
applications in the context of elliptical galaxies, see Marchant \& 
Olson (1979), Richstone (1979), and Olson \& de Vaucouleurs (1981).] 

In Figures 2 and 3 we plot the relation between log $I$ and log $p$ for 
each subsample of the combined continuum dataset. The sample of Chini et 
al.\ (1997) is shown in Figure 2$a$. This sample contains only 21 data 
points, and while there is some suggestion of an increasing trend of 
$I$ with $p$, it is not statistically significant. The same is true of 
the Motte et al.\ (1998) data from Ophiuchus, shown in Figure 2$b$. 

While the above two samples show no significant trend, the Orion B 
dataset of M01 ($N = 64$ resolved objects), shown in Figure 3$a$, is 
more definitive. Although there is considerable
scatter, a correlation does exist between log $I$ and log $p$. 
The linear (parametric) correlation coefficient (C.C.) is 0.35, with 
a significance of greater than 99 percent. Under the more general 
assumption that the underlying distributions of $I$ and $p$ are 
not binormal, we can perform the nonparametric Spearman and Kendall 
(rank correlation) tests. These give a coefficient of 0.29 and a 
significance of 98 percent. 
The point at (log $p$, log $I$) $\simeq (-0.05, 4)$ is clearly an 
outlier. It corresponds to NGC 2071--IRS, a known outflow source. 
Since we wish to focus on starless cores in this study, we exclude 
it from further consideration.\fn{In doing so we err on the side of 
caution, since NGC 2071--IRS is still deeply embedded, with an extended 
envelope whose shape may still reflect the pre-stellar condition of 
the core.} This lowers the rank C.C. to 
0.27 with a significance of 96.3 percent. The best-fit straight line 
to the data is given by  log $I = A~{\rm log}~p + B$,
with $A = 0.50 \pm 0.18$ and $B = 2.56 \pm 0.06$. 

The above fit was 
performed neglecting the stated observational errors in log $I$ and 
log $p$ (\S \ref{sec-data}). While this might be considered reprehensible, 
we found that including the errors gave a goodness-of-fit parameter 
that was too low to be acceptable by the usual standard 
($\gae 10^{-3}$; Press et al.\ 1992). 
This result does not cast aspersions on the validity 
of the correlation stated above (which does not depend on the errors), 
nor need it necessarily lower our confidence in the fitted values of 
$A$ and $B$. Rather, it is a familiar 
feature of data that have a scatter larger than the formal errors. In 
particular, it indicates that the uncertainties ascribed to $I$  
are {\it underestimates.} We present ample support for this statement, 
along with a consistency check of the above fitted parameters, in \S 
\ref{sec-simul}. There a more robust analysis is used to determine the 
model goodness-of-fit for the data (specifically, the Monte Carlo 
simulation of synthetic data sets). 

Many of the cores (18/82, or 22 percent) in the sample are
unresolved, and therefore are not included in the above fit. 
Interestingly, despite the fact that all are 5$\sigma$ detections, 
none has a peak intensity exceeding 0.25 Jy/beam. If the apparent 
ellipticities  
of the unresolved cores are distributed in the same way as the 
rest of the sample, this then suggests that there are more elongated 
than round sources amongst this sub-population. 

Finally, we note that while the observed slope of the above 
fit is positive, it disagrees with the value of $+1$ expected if all 
of the cores are randomly-oriented, prolate spheroids with $I_p =$ const. 
This feature will be addressed in \S \ref{sec-intrins}. 

\subsection{Relation Between Total Flux and Apparent Ellipticity}

The observed intensity integrated over the core area, or total flux
density, $S$, is a quantity of additional interest, since it is 
proportional to the 
core mass $M$. Again assuming the dust emission to be optically 
thin, the gas mass of a core is (Johnstone et al.\ 2000)
\bed
M = \frac{S~d^2}{\kappa~B (T_d)},
\eed
where $d$ is the core distance, $T_d$ the dust temperature, $B$ 
the Planck function, and $\kappa$ the dust opacity (the latter two 
evaluated at a specific frequency of interest). 
In the absence of any intrinsic dependence of $S$ on $q$, it is clear 
that $S$ {\it should be independent of $p$, since the emission is 
optically thin.} This furnishes an important test of the hypothesis
that the observations can be reproduced by a population of 
more or less identical (i.e., $I_p \approx$ constant), randomly-oriented 
cores. 

A log-log plot of $S$ vs.\ $p$ for the M01 sample is shown in Figure 
3$b$. It shows that $S$ exhibits a strong positive correlation with 
$p$. Again omitting NGC 2071-IRS, the linear correlation coefficient 
is 0.37, with a significance of greater than 99 percent. The rank 
correlation is 0.31 with a significance of 98.7 percent. The best-fit 
straight line to the data has $A = 0.65 \pm 0.21$ and $B = 2.84 \pm 0.06$. 
Thus it appears that {\it intrinsic effects do play a role in
producing the observed peak intensity and mass distribution of the cores.} 
The dependence of $I$ and $M$ on $q$ implied by this result will be examined
in \S \ref{sec-intrins}. 

Another interesting property of the total flux is the 
degree to which it tracks the peak intensity $I$. A log--log plot of 
$S$ versus $I$ yields a tight correlation with a rank C.C.\ of 0.89 
(probability of no correlation $< 10^{-4}$) and a slope of $1.17 \pm 
0.06$. This is 
not that surprising given that the mean core size $\lb R \rb \simeq 
5500$ AU just exceeds the beam size of 5000 AU. However, it is easier 
to reconcile with an intrinsically prolate distribution in the following 
respect. Members of the latter population, having $b' = b$, will more 
often be viewed near the resolution limit, since $\lb b' \rb = 4600 
\pm 2900$ AU. In the oblate case, however, $\lb a' \rb = 7800 \pm
4500$ AU, so that nearly all objects are greater than the beam size 
in extent, leading to a bigger expected difference between $S$ 
and $I$. 

Finally, by plotting log $S$ versus the log of the core size $R 
\equiv (a'b')^{1/2}$,  
we find---confirming the conclusion of M01---that $S \propto
R^{1.1}$. This is close to the $M \propto R$ relation expected for 
a gaseous sphere in virial equilibrium, suggesting that the cores 
are self-gravitating. 
 
\section{INTRINSIC CORRELATIONS}
\label{sec-intrins}
\subsection{General Considerations}
\label{sec-consid}
The results of the previous section imply that $I$ depends upon 
intrinsic properties of the core. 
We first ask whether there exist any systematic trends that  
could contribute to the correlations seen in Figure 3.
For instance, were the size of the cores to increase 
with increasing $q$---irrespective of any assumption about intrinsic 
shape---then both $I$ and $S$ would be seen to increase with $p$. 

To look for such an effect, we plotted 
log $I$ vs.\ log $R$ and the observed FWHM major axis log $a'$ 
vs.\ log $p$ in Figures 4$a$ and $b$, respectively. In the first plot,
no correlation is seen---although the highest flux point, NGC2071-IRS, 
has one of the largest inferred sizes. A curious feature 
of the second plot is the exactly linear slope of $-1$ seen in the 
subset of points at lower left. This is a resolution effect: 
all nine objects have a projected minor axis $b' = 1600$ AU, the 
smallest in the sample. Since $p = b'/a'$, each point lies 
on a single line of slope $-1$ in the log $a'$--log $p$ plane. Several 
such alignments can be seen in the plot, reflecting the finite spatial 
resolution of 200 AU. Statistically, however, there is no significant 
correlation between the two quantities. Finally, the observed size
distribution of the cores is close to gaussian, with $\lb$log $R \rb 
= 3.74 \pm 0.17$. 

In the absence of a systematic variation in scale with $p$, there 
remains the possibility of an intrinsic dependence of $I_p$ on 
$q$ [equations (\ref{eq-Sob}) and (\ref{eq-Spl})]. In the context of 
elliptical galaxies, Merritt (1982) showed that such a dependence can 
completely mask or even reverse the expected correlation in (log $I$, 
log $p$). Following his analysis, we begin by assuming that $I_p$ 
depends only on $q$, and write 
\beq
I_p = q^m I_0,
\label{eq-Spofq} 
\eeq
where $I_0 =$ constant and $m \neq 0$ (the discussion 
of the previous section assumed $m=0$). As Merritt demonstrated, a 
positive slope of unity in (log $I$, log $p$) may be obtained either
by assuming that all objects are prolate with $m$ near zero, or that 
all objects are oblate with $m$ large and positive.  
Further, once observational errors were added, simulated data 
corresponding to, say, a prolate population with $m=0$ and an 
oblate population with $m=3$ became formally 
indistinguishable.\fn{Note that Merritt adopted a different 
convention for $p$ from that used here ($p < 1$ in the oblate case; 
$p > 1$ for prolate), used a different intrinsic distribution 
$\psi (q)$, and plotted $|{\rm log}~p|$ on the horizontal axis. 
Hence, the signs and values of his slopes generally differ from ours.}
It is desirable to determine what range of $m$ values may be expected 
in the context of molecular clouds. Here we turn to theoretical 
considerations which, however crude, can be used to obtain rough 
constraints on this parameter.  

\subsection{Oblate clouds}
\label{sec-oblate}
Several studies of isolated oblate clouds---the 
prevailing theoretical paradigm for the precursors of 
protostars---have established the relation between ellipticity and 
mass: in general, both increase together. This is true irrespective
of whether the flattening is caused by rotation (Stahler 1983;
Kiguchi et al.\ 1987) or flow down polar magnetic field lines
(Tomisaka, Ikeuchi, \& Nakamura 1988; 
Nakamura, Hanawa, \& Nakano 1995), provided that only those 
equilibrium sequences along which the equatorial radius is roughly 
constant are considered.\fn{In the magnetic case, such a 
sequence has a constant ratio of magnetic pressure at infinity to 
cloud surface pressure. In the 
rotating models, the ratio of the rotational to gravitational energy at 
the cloud surface is held fixed. Both of these assumptions are  
reasonable if members of the sequence are taken to represent  
different stages of condensation---within a single parent cloud---in 
the approach to dynamical collapse.} 
It is the tendency of gravity to enhance elongation along the shorter 
axis (i.e., $q \propto M^{-\alpha},~ \alpha > 0$) that is responsible
for the flattening in both cases.

In order to relate $M$ to the assumed form of $I_p$, first recall 
that $S \simeq I^{1.17}$ in the M01 sample. In general, we expect 
$S \sim I^{1+C}$, where $C > 0$ is a constant. Then, 
since $M \propto S$, we have
\beq
M \sim I^{1+C} = (p^{-1}I_p)^{1+C} = (p^{-1} q^m I_0)^{1+C}.
\label{eq-obmass}
\eeq
According to equation (\ref{eq-obltheta}), $p$ ranges from $q$ to
unity depending upon orientation, so $M \sim q^{(m-1)(1+C)}$ if all 
cores are observed edge-on and $M \sim q^{m(1+C)}$ if all are pole-on. 
In order for $M$ to increase with decreasing $q$ then, {\it we must at 
the very least have $m < 1$ in the oblate case.} 
This result agrees qualitatively with that obtained in the special 
case of a uniform, isothermal (sound speed $=c_s$), self-gravitating 
disk (Fleck 1992):
\bed
M \simeq \frac{c_s^2~a}{2G}~q^{-1}, 
\eed
which corresponds to $\alpha = 1$ in the above canonical form 
(recall that $a$ is the major axis length). Larger or smaller values 
of $\alpha$ might apply to less flattened configurations, a non-isothermal 
equation of state, and/or an inhomogeneous matter distribution. 
In any event, it would appear that {\it in the oblate case, the 
large positive values of $m$ required to produce a positive slope 
in} (log $I$, log $p$) {\it are effectively ruled out.} 

\subsection{Prolate clouds}
\label{sec-prolate}
The lack of a theoretical paradigm for prolate clouds precludes a 
similarly firm prediction in this case. However, from the few studies 
that exist (Hanawa et al.\ 1993; Curry 2000; Fiege \& Pudritz 2000a; 
Balsara et al.\ 2001), certain key properties may be anticipated. First, 
prolate clouds are not likely to be isolated, but rather should be embedded 
in larger, filamentary structures.\fn{While isolated prolate equilibria 
have been constructed using specific magnetic field geometries (Fiege 
\& Pudritz 2000b; Curry \& Stahler 2001), these can hardly be considered 
generic.} Second, the behavior of an embedded core's mass and density 
as a function of its intrinsic axis ratio is such that the latter rises 
to near unity at high mass and central concentration (Curry 2000;
Fiege \& Pudritz 2000b). Thus, a prolate core condensing out of its
parent filament should  
approach an approximately spherical shape at large mass, or 
equivalently, at a late evolutionary stage as a pre-stellar object. 

The analogous expression to equation (\ref{eq-obmass}) in the prolate 
case is 
\beq
M \sim I^{1+C} = (p~I_p)^{1+C} = (p~q^m I_0)^{1+C}. 
\label{eq-plmass}
\eeq
With $p$ again ranging between $q$ and unity, this gives 
$M \sim q^{(m+1)(1+C)}$ if all the cores are edge-on and 
$M \sim q^{m(1+C)}$ 
if all are pole-on. In order for $M$ to increase with increasing $q$
then, {\it we must at the very least have $m > -1$ in the prolate case.} 

\subsection{Consequences for the observed correlations}
\label{sec-conseq}
These predictions for the expected range of $m$ in the two cases should 
be taken as rough guidelines; exceptions can surely be found. However, 
to the extent that they are reasonably robust, they lead to the following 
conclusions. Using equations (\ref{eq-Sob}), (\ref{eq-Spl}), and 
(\ref{eq-Spofq}), we write the log $I$--log $p$ relation as
\bed
{\rm log}~I = \pm~{\rm log}~p + m~{\rm log}~q + {\rm log}~I_0,
\eed
where the upper (lower) sign on the right-hand side applies to the 
prolate (oblate) case. In the oblate case, it is clear that the effect 
of an intrinsic dependence of $I_p$ on $q$ with $m < 0$ is an 
{\it increase} in the mean slope of the log $I$--log $p$ relation from 
its $m=0$ value of $-1$. A decrease in slope can only occur if 
$0 < m < 1$; however, the slope is still limited to negative values,  
since $q \leq p$. In the prolate case, on the other hand, the 
expected dependence of $I_p$ on $q$ leads to a {\it decrease} in the 
slope if $-1 < m < 0$, and an {\it increase} if $m > 0$.
This suggests that small negative values of $m$ may be consistent 
with the observed slope of the (log $I$, log $p$) relation. We test 
this hypothesis in the following subsection.

\subsection{Simulations}
\label{sec-simul}
Using the relations found in this and the previous sections, Monte
Carlo simulations were carried out for a sample of 65 model cores. 
Each simulation consisted of 1000 realizations of a given model. 
In detail, the procedure was as follows. 
(1) Each core was randomly assigned an intrinsic ellipticity from the 
best-fit gaussian distribution (either oblate or prolate) derived in 
\S \ref{sec-Nvsp}. (2) The angle of observation was randomly chosen 
from a uniform distribution over the sphere. (3) The intrinsic flux 
log $I_0$ was, in the first instance, set equal to 2.45, the mean of 
the observed dataset. (4) Normal distributions of measurement 
error were added randomly to both $p$ and $I$. In accord with the 
uncertainties cited in M01 and \S \ref{sec-data}, we assigned 
dispersions of 0.05 in log $p$ and 0.15 in log $I$. (5) In order 
to test the hypothesis that some of the scatter in the observational 
plot might be intrinsic to the cores, we also ran simulations in which  
an ad hoc variation of $\pm 0.15$ in log $I_0$ was added to some of 
the realizations. This intrinsic scatter could, e.g., reflect a dependence 
on an unknown parameter, or simply be due to random differences in 
formation or past history of individual objects. 
(6) Finally, using equations (\ref{eq-obmass}) and (\ref{eq-plmass}), 
we calculated $S$ for each simulated core. 

The results are given in Tables 1 (log $I$ vs.\ log $p$) and 2 
(log $S$ vs.\ log $p$). Calculations were performed 
for values of $m$ appropriate to the ranges derived in the previous
subsections. For each model there are two entries: the mean slope and the 
rank correlation coefficient, averaged over all 1000 realizations of the 
model, and their dispersions. A model was 
judged a good fit to the observations at the $1\sigma$ level if the 
observed values of both the slope and correlation coefficient (Figure
3$a$) fell within the $1\sigma$ confidence intervals of the Monte Carlo 
simulations. 

As expected from the results of the preceeding subsection, the oblate 
model slopes in the (log $I$, log $p$) plane are negative in the range 
examined, while prolate models have positive slopes (see Table 1). 
Evidently, introducing an intrinsic scatter in $I_0$ hardly alters 
the derived slopes; only the dispersions are affected. Not surprisingly, 
the correlation coefficients are also lowered in models with intrinsic 
scatter, particularly near $m = 0$ (oblate) and $m = -1$ (prolate). 
Note that the simulations correctly reproduce (within the dispersions) 
the expected slopes for $m=0$; namely $-1$ and $+1$ in the oblate and 
prolate cases, respectively. 

Comparing now with the observed correlation in log $I$--log $p$ 
(slope = 0.50 $\pm$ 0.18, rank C.C. = 0.27; Figure 3$a$), we see that 
a prolate model with $-1 < m < -0.5$ is likely to reproduce the 
observed parameters. Specifically, it is found that for $m = -0.60$ 
and $\sigma$(log $I_0$) = 0.15, the simulated slope and C.C. are 
0.50 $\pm$ 0.15 and 0.26 $\pm$ 0.08, respectively. Therefore, this 
set of prolate models is in agreement with the observations at the 
$1\sigma$ level. One realization from this suite of simulations is 
shown in Figure 5$a$. 

The $S$ vs.\ $p$ results given in Table 2 provide an important, 
although not entirely independent, consistency check of these
results. The above best-fit simulation with $\sigma$(log $I_0$) = 
0.15 and $m = -0.60$ has a slope (in the log $S$--log $p$ plane) 
of 0.59 $\pm$ 0.16 and a rank C.C.\ of 0.28 $\pm$ 0.08, respectively. 
While a better fit to the data of Figure 3$b$ could be found for a 
slightly larger value of $m$, we do not feel that this would be as 
reliable a determination as that found above, 
given the uncertainties in the $S(q)$ relations derived in \S 
\ref{sec-consid}. A realization from this suite of simulations is 
shown in Figure 5$b$. 

Recall that, in \S \ref{sec-Svsp}, the linear least-squares fit 
{\it with errors} to the observed (log $I$, log $p$) gave a 
goodness-of-fit parameter that was too low to be acceptable.  
It is worth checking whether the addition of the intrinsic scatter 
in $I_0$ rectifies this situation. Upon doing so, we found a slope 
of $0.59 \pm 0.20$ and
intercept of $2.60 \pm 0.06$, with a goodness-of-fit equal to 0.41. 
These values are in agreement with those found ignoring the errors 
in \S \ref{sec-Svsp}, thus providing an extra measure of confidence 
in the fitting parameters derived there. 

Finally, it is interesting to ask whether a satisfactory fit of the oblate
models to observations could be obtained if $m$ were unrestricted. 
Surprisingly, the answer is no: while a slope of $0.50 \pm 
0.88$ in (log $I$,log $p$) is obtained for $m = 4.2$ in the oblate case, 
the corresponding rank C.C.\ 
equals 0.02 $\pm$ 0.08, indicating no correlation. This means that the 
underlying distribution of axis ratios $\psi (q)$ prohibits such a 
correlation. Thus, we conclude that the oblate hypothesis can be rejected 
with the same degree of confidence.  

\section{DISCUSSION}
\label{sec-discuss}
\subsection{Intrinsic Correlations: Theory}
\label{sec-theory}
Here we compare the results obtained above with a particular class  
of prolate equilibria: the embedded, isothermal cores of Curry (2000). 
The equilibrium sequence is characterized by a single parameter, $2Z$: 
the length of the parent cylindrical cloud, as compared with the critical 
wavelength for instability along its major axis, $\lambda_{\rm cr}$. 
The embedded condensations exist only for $2Z > \lambda_{\rm cr}$. 
In standard cylindrical coordinates, the 
point $r=0,~z=Z$ represents a saddle point between two adjacent cores. The 
cores themselves are defined by the tidal lobe, the constant-density surface  
extending from $(0,Z)$ to $(0,-Z)$ through the point $(R_{\rm tl},0)$ in 
the midplane (see Figure 2 of Curry 2000). It is desirable to 
calculate the variation of both the polar intensity (or column density) and 
the mass of the embedded cores as a function of intrinsic ellipticity in 
these models. 

We are immediately confronted with the issue of how to define the 
above quantities in the context of an embedded core. Specifically, 
what are the appropriate theoretical definitions of $I_p$ and $M$? 
The former is proportional to the column density along the core 
major axis,
\beq
N_p = \int_{-Z}^Z \rho(0,z)~dz. 
\label{eq-zcol}
\eeq
As $Z \goto \lambda_{\rm cr}/2$, $\rho (r,z)$ approaches $\rho_{\rm 1D}(r)$, 
the density profile of a 1D, isothermal filament (Stodolkiewicz 1963; 
Ostriker 1964). Thus, the quantity given by 
equation (\ref{eq-zcol}) approaches $N_p = \lambda_{\rm cr} \rho_{\rm 1D} 
(0)$ in that limit, declining to zero for smaller $Z$. At larger $Z$, 
the isodensity contours are prolate (inside the tidal lobe; outside, they 
have an open topology), and approach a spherical shape at the largest 
$Z$. The latter equilibria are the most centrally-concentrated 
of the sequence.

Regarding the corresponding behavior of the mass, we first calculated 
the isodensity surface corresponding to the half-maximum of the column 
density; i.e., the contour corresponding to $N_p/2$. Let the polar and 
equatorial radii of the enclosed volume be denoted by $z_{1/2}$ and 
$r_{1/2}$, respectively. The bounding surface is the 
analogue to the FWHM intensity contour of the observations. The 
enclosed mass, $M_{1/2}$, can then be found by direct integration. 

Figure 6 shows the behavior of $N_p$ and $M_{1/2}$ as a function 
of $q (= r_{1/2}/z_{1/2})$ in a log-log plot. The polar column density 
is remarkably flat over the range of 
$q$ examined, while $M_{1/2}$ is an increasing function of $q$, as 
expected.\fn{Note that, due to the different choice of bounding surface, 
the behavior of $M_{1/2}$ is qualitatively different from that of 
$M_{\rm tl}$, which has a peak at intermediate $q$ (Curry 2000).} The 
dotted line shows the $I_p \propto q^{-0.6}$ best-fit dependence derived 
from the Monte Carlo simulations. It is reassuring that there is 
qualitative agreement of $N_p$ with this line for $q \gae 0.25$. The 
behavior of $M_{1/2}$ is also qualitatively consistent with expectation, 
as it has a positive slope ($\simeq 1.25$) toward $q=1$. However, the 
latter slope disagrees quantitatively with the prediction of equation 
(\ref{eq-plmass}) above, which gives a maximum slope of $(m + 1)(1 + C) 
= 0.40(1.17) = 0.47$ for the M01 sample. 

Hence it appears that the results of the shape analysis can be reconciled, 
at least qualitatively, with one theoretical model of prolate cores. As the 
latter is undoubtedly among the most simple one can construct, it remains to 
be seen whether more realistic models can provide an equal or even superior 
level of agreement. 

\subsection{Non-Random Orientation}
\label{sec-orient}

Here we examine the consequences of relaxing the other principal assumption 
of this work: that of random orientation. Dense cores are nearly always 
found embedded within larger molecular clouds which, in turn, 
often display a filamentary appearance. For example, in a recent study 
of Ophiuchus, the major axes of cores detected in C$^{18}$O were 
found to correlate with the symmetry axes of the larger filaments,
detected in $^{13}$CO (Tachihara et al.\ 2000). Thus, it may be 
that all cores within a given filament have roughly the same orientation, 
$\theta_0$, by dint of their embedding. Then, from equations 
(\ref{eq-obltheta}) and (\ref{eq-protheta}), $p$ is given by 
\beqa
p(q) &=& [1 - (1 - q^2)~{\rm cos}^2 \theta_0]^{1/2} ~~~~~~~~~~~~~
({\rm oblate}), \label{eq-obtheta} \\
 &=& q~[1 - (1 - q^2)~{\rm sin}^2 \theta_0]^{-1/2} ~~~~~~~~~
({\rm prolate}). \label{eq-pltheta} 
\eeqa

Now imagine that there exists a distribution of intrinsic axis ratios 
$\psi (q)$ that is non-zero at each $q$.
In the oblate case, this implies {\it a minimum, non-zero value of $p$ 
in the observed distribution of axis ratios}, and so a reduced range of 
possible $p$ (e.g., a lack of edge-on objects unless 
$\theta_0$ were exactly equal to zero). Indeed, since the median value 
of $\theta_0 = \pi/3$ corresponds to a minimum $p = 0.87$, it is unlikely 
that even an intrinsically very flattened object would appear elongated. 
Conversely, a distribution of oblate cores with only a very few 
intrinsically round objects is able to reproduce the observed $\phi (p)$ 
near $p = 1$. While the observed distribution of $p$ does display 
a preponderance of values near $p=1$, there is no clear
cutoff at lower $p$ (Figure 1). 

On the other hand, equation (\ref{eq-pltheta}) shows that a distribution 
of prolate objects still produces the 
entire possible range of $p$. Only for $\theta_0$ very near $\pi/2$ 
does $p(q)$ rise to near unity at small values of $q$. Interestingly, 
this means that unless the intrinsic distribution $\psi (q)$ contains 
a significant number of (intrinsically) nearly round objects, the 
observed $\phi (p)$ will be deficient in (apparently) round objects 
unless $\theta_0 \simeq \pi/2$. Moreover, extremely elongated
prolate objects should be quite faint (i.e., difficult to distinguish 
from the background continuum, due to their comparatively low column 
density), so 
any observational sample will have a cutoff at some minimum value 
$q_\mn$, with corresponding apparent ellipticity $p_\mn \geq q_\mn$. 
For any $\theta_0 > 0$ then, the range of $p$ will be further reduced. 
The presence of relatively few objects 
with small $p$, and therefore small $I$, could lead to an artificial 
weighting of higher$-I$ objects and therefore a shallower slope than 
$+1$ in log $I$ vs.\ log $p$, as is observed.  
A more detailed examination of inclination effects, with a corresponding 
analysis of the observed position angle distribution of cores, would be 
a highly worthwhile undertaking.

\subsection{Implications for Core Formation in Filaments}

As to the implications of these results for star formation generally,  
a new picture is emerging of pre-stellar condensation occuring 
predominantly in filaments. Thus, at least initially, the 
condensations inherit the shape of their surrounding filament, 
while gaining more mass along the filament axis. As the core 
grows in mass, its self-gravity pulls it into a more spherical 
shape, at the same time causing it to detach from the parent cloud. 
Only after this point might the previous paradigm of isolated, 
low-mass star formation become relevant (Shu, Adams, \& Lizano 1987). 

Aspects of this picture have been touched upon by previous authors. 
Schneider \& Elmegreen (1979), motivated by the appearance of ``globular 
filaments'' in optical extinction maps, proposed that gravitational 
instability in the filaments produces the denser, embedded globules.
In some cases, the latter were approximately equally spaced along the 
background filament axis (see also Dutrey et al.\ 1991). 
Fleck (1992) emphasized that gravity acting on a nonisotropic mass 
distribution could be responsible for some aspects of the observed 
shape distribution. However, he claimed that this implied 
``molecular clouds are not generally in global (three-dimensional) 
equilibrium.'' Further, he argued that, independent of the exact 
nature of core elongation (e.g., oblate vs.\ prolate spheroidal), 
smaller mass objects should be more nearly spherical ``and, because 
of their low self-gravity, shaped by forces other than gravity.'' 
These comments echo those of earlier authors (Lin, Mestel, \& Shu 1965; 
Larson 1985; Bastien et al.\ 1991), who remarked that prolate 
configurations might result from the dynamical contraction of initially 
cylindrical clouds. 

However, it is clear that these dynamical scenarios bear little resemblance 
to the embedded states considered above. First, as a function of increasing 
core condensation and mass, the intrinsic ellipticity behaves in an opposite 
manner in the two cases. Second, the embedded 
states originate from the spontaneous fragmentation of an initial filament 
which is near marginal stability, {\it but not dynamically unstable} 
(Curry 2000). Third, and most obviously, observations do not furnish any 
examples of filaments with true cylindrical symmetry. Rather, filamentary 
structure is invariably a characteristic of cloud envelopes, with 
the core regions displaying significant fragmentation. Finally, we 
draw attention to a specific mechanism for embedded core formation 
presented by Balsara et al.\ (2001). In this scheme, matter is channelled 
onto a growing core by a magnetic field more or less coincident with the 
filament axis. Although the accretion occurs on a dynamical timescale in 
this scenario, it nevertheless gives some indication that these simple 
theoretical ideas can be extended into more complex physical regimes. 
 
\section{CONCLUSIONS}
\label{sec-conc}
Previous studies of dense core morphology, nearly all of which used 
molecular line data, gave varied results as to the intrinsic shapes 
of the cores. We have employed recent dust continuum datasets and 
additional methods of analysis in an effort to clarify the situation. 
The main conclusions of this work are as follows: 
\vskip 0.1 cm
\noi
(1) The observed distribution of core ellipticities in the combined 
continuum sample is well fit by a gaussian distribution of intrinsically 
prolate objects with mean ellipticity $\lb q \rb \approx 0.5 \pm 0.2$. 
\vskip 0.1 cm
\noi
(2) In the M01 sample, the peak intensity $I$ is positively 
correlated with the apparent ellipticity $p$, with a slope in 
log $I$ vs.\ log $p$ of $0.50 \pm 0.18$. This slope is shallower 
than the $+1$ value expected for an ensemble of randomly-oriented  
prolate spheroids, each having constant polar intensity $I_p$. 
\vskip 0.1 cm
\noi
(3) In the same sample, an equally significant correlation (slope $= 
0.65 \pm 0.21$) is observed between the log of the total flux 
density $S$ and log $p$. This shows that both $I_p$ and the mass of 
a given core depend on its intrinsic ellipticity, $q$.
\vskip 0.1 cm
\noi
(4) Under the assumption that $I_p = q^m I_0,~I_0 =$ constant,  
Monte Carlo simulations were used to find the value of $m$---and 
the spheroidal shape---that best fits the observed correlation in 
(log $I$, log $p$). The observed slope and rank C.C.\ are best fit 
by a prolate ensemble with $m = -0.60$ and an intrinsic scatter 
in log $I_0$ = 0.15. No satisfactory fits for oblate spheroids 
were found. 
\vskip 0.1 cm
\noi
(5) The relation $I_p = q^{-0.6} S_0$ was shown to agree (for $q \gae 
0.25$) with the expected polar column density in the embedded prolate 
equilibrium sequence of Curry (2000). 
\\

The chief limitation of the present work is the relatively small 
number of objects in the M01 sample, and the consequent weakening 
of the statistical results so obtained. Also, the fact that all of the
cores come from mainly filamentary structures found in three regions 
may be considered a bias. However, if these cores are truly
pre-stellar (as suggested by their number distribution as a function 
of mass; see \S \ref{sec-intro}), then this mode of 
condensation may in fact be reasonably representative of low-- 
and intermediate--mass star formation. The possibility that at least 
some of these highly embedded objects are aligned with the major 
axes of their parent filaments is real, and may alter the 
results presented here in certain respects (\S \ref{sec-orient}).   
The addition of new dust continuum data will allow more definitive 
conclusions to be drawn on each of these key points, and will no doubt 
aid in the formulation of more sophisticated theoretical models of 
embedded cloud equilibria.  
\\\\
\noi
I am grateful to F.\ Motte for providing her previously unpublished 
peak fluxes, and thank Carol Jones for preliminary discussions about 
this work. Chris McKee, Steve Stahler, and an anonymous referee
offered valuable comments that helped improve the original manuscript.

\clearpage

\bfig
\epsscale{0.55}
\plotone{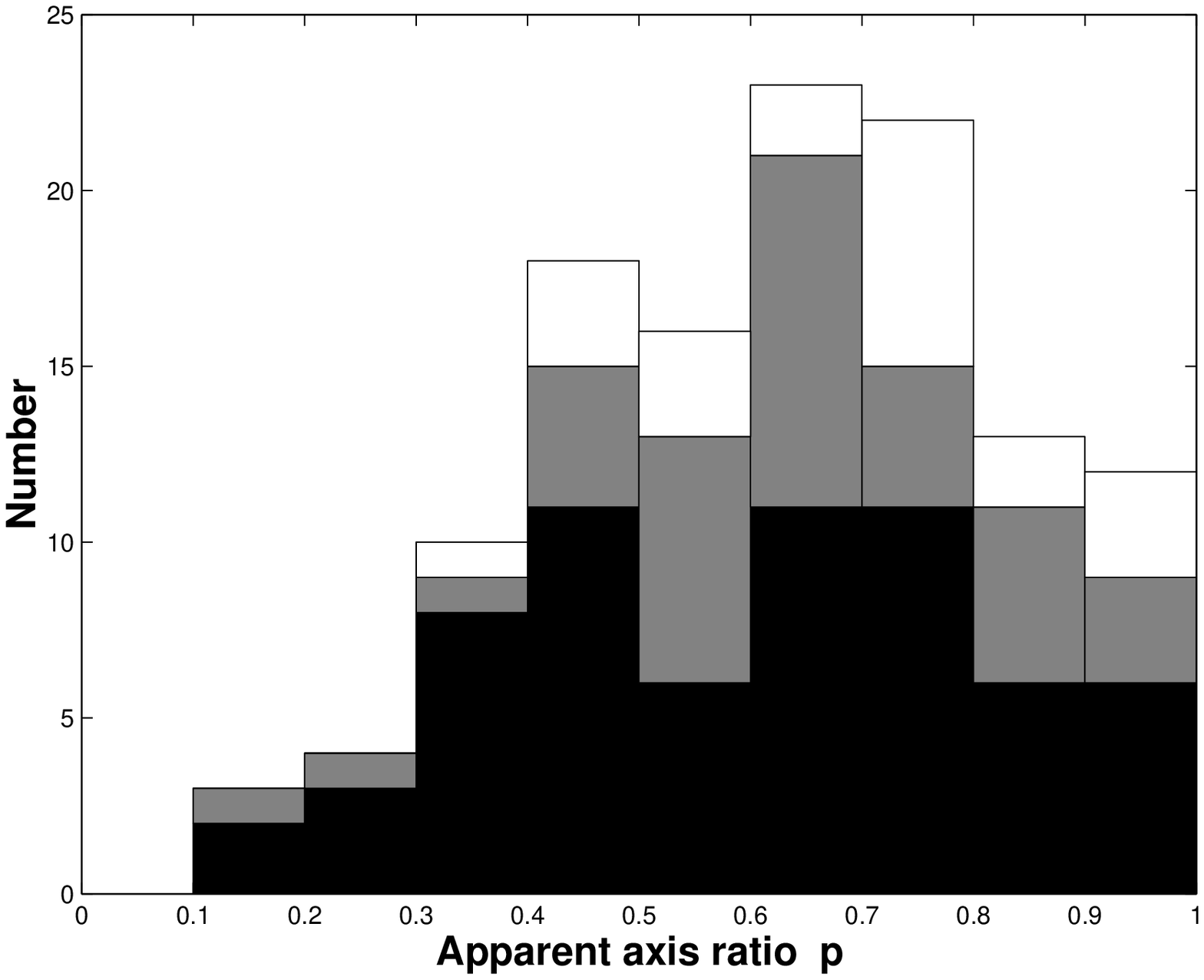}
\vspace{1cm}
\epsscale{0.55}
\plotone{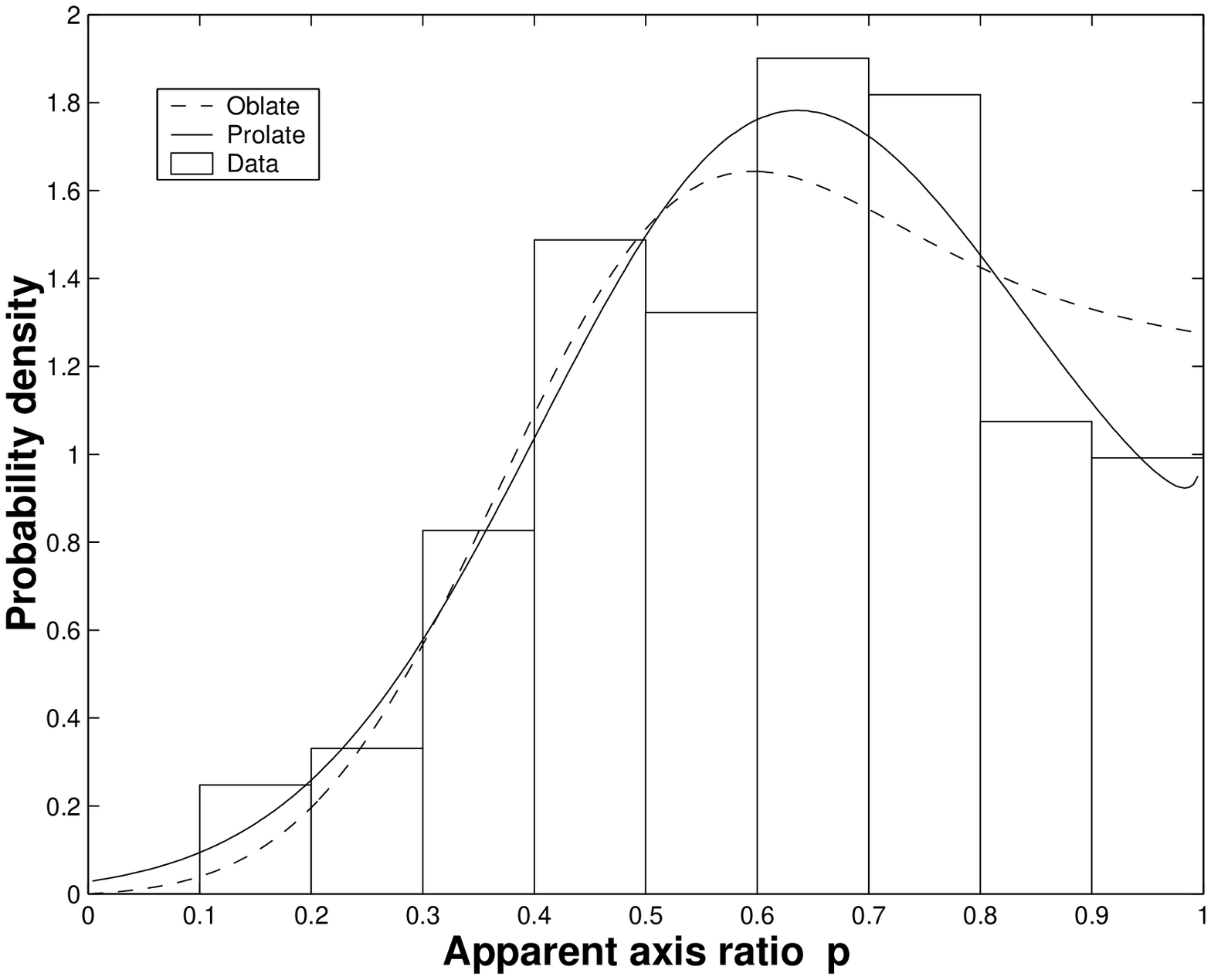}
\caption{(a) The number distribution of apparent axis ratio in the
combined continuum sample. The portions belonging to individual 
samples are indicated in each bin: Motte et al.\ (2001) {\it (black)}; 
Motte et al.\ (1998) {\it (gray)}; Chini et al.\ (1997) {\it (white)}. 
The histogram contains 121 cores in total. 
(b) The best-fit oblate (dashed) and prolate (solid) gaussian 
distributions to the observed histogram, normalized as a
probability density. See text for the best-fit parameters.}
\efig

\bfig
\epsscale{0.55}
\plotone{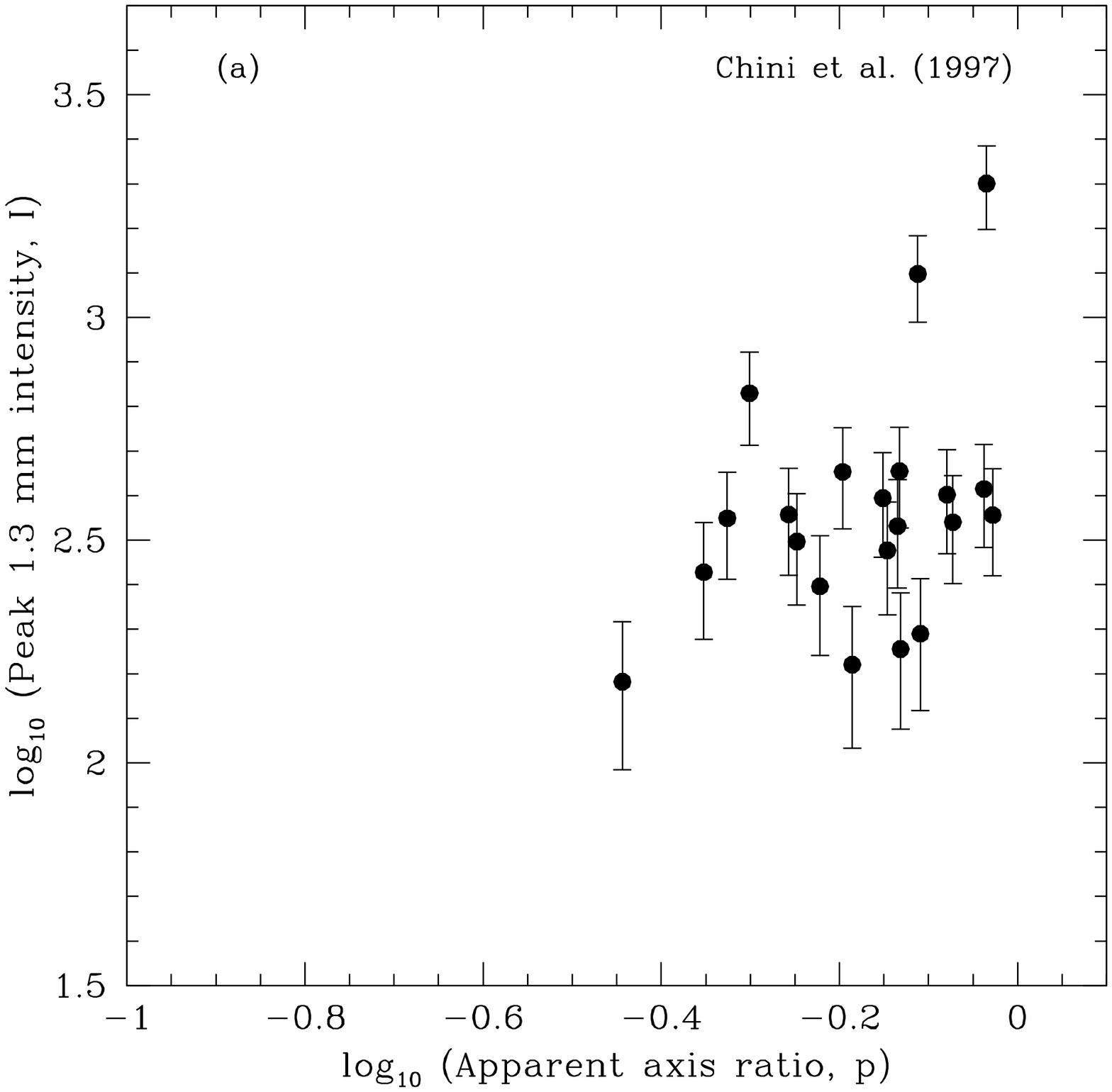}
\vspace{1cm}
\epsscale{0.55}
\plotone{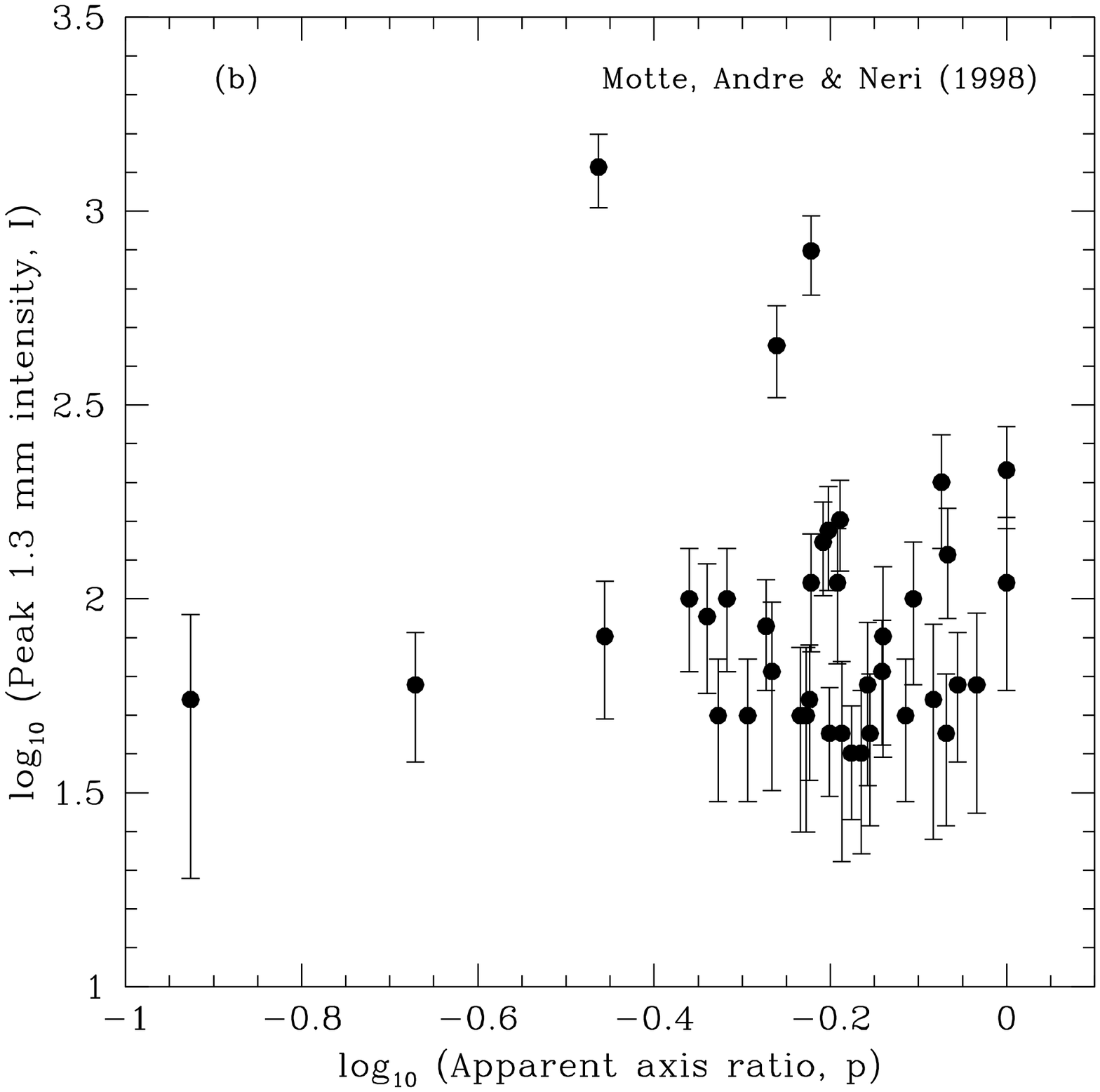}
\caption{(a) Peak intensity as a function of apparent ellipticity 
for the sample of Chini et al.\ (1997). (b) Same as (a), but for the 
sample of Motte et al.\ (1998). In both samples, the error bars reflect 
the calibration uncertainty of 20\% plus the rms noise cited in the 
respective papers. Approximate errors in log $p$ (omitted for 
clarity) are $\pm 0.07$ in (a) and $\pm 0.05$ in (b).} 
\efig

\bfig
\epsscale{0.55}
\plotone{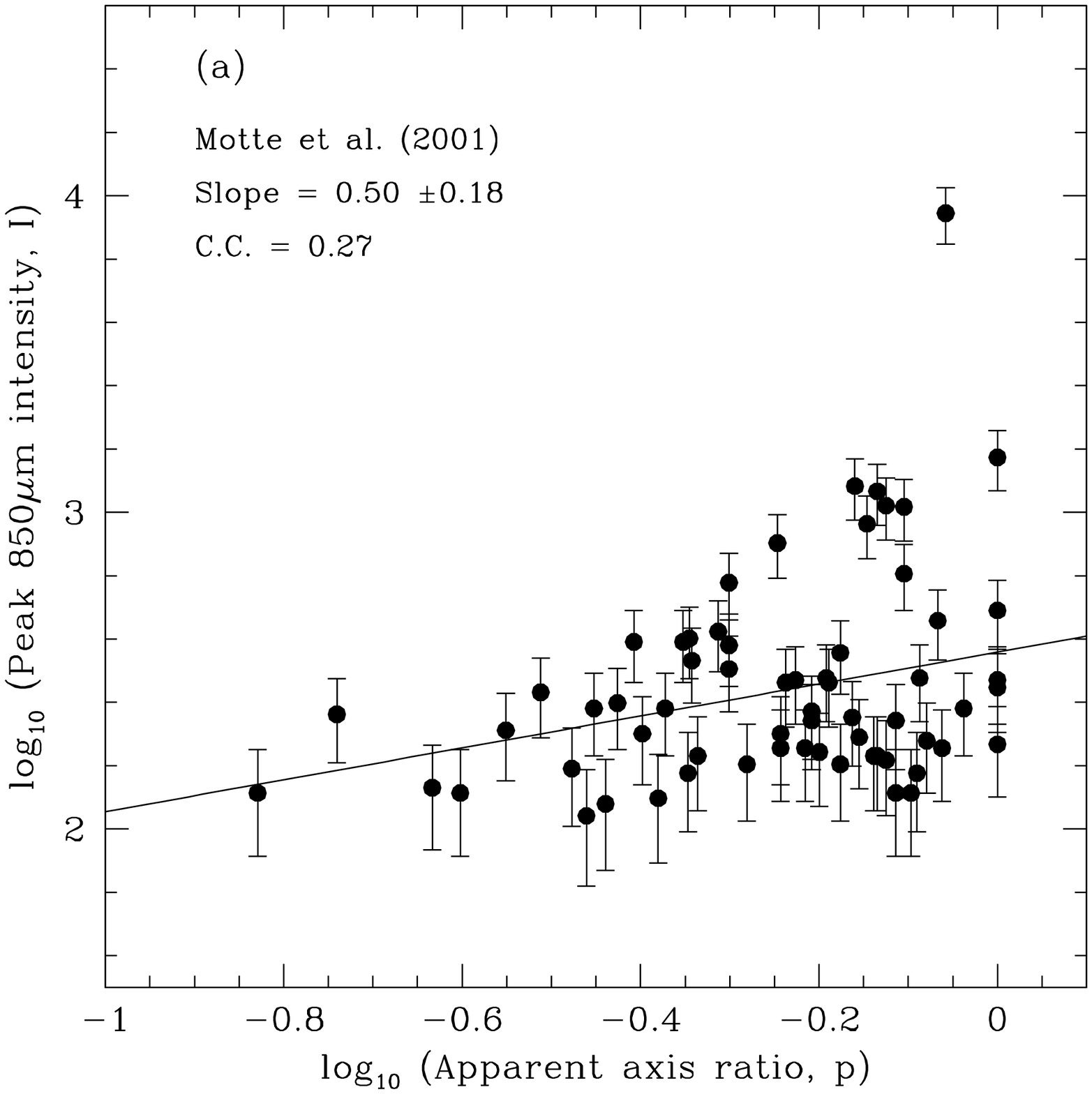}
\vspace{1cm}
\epsscale{0.55}
\plotone{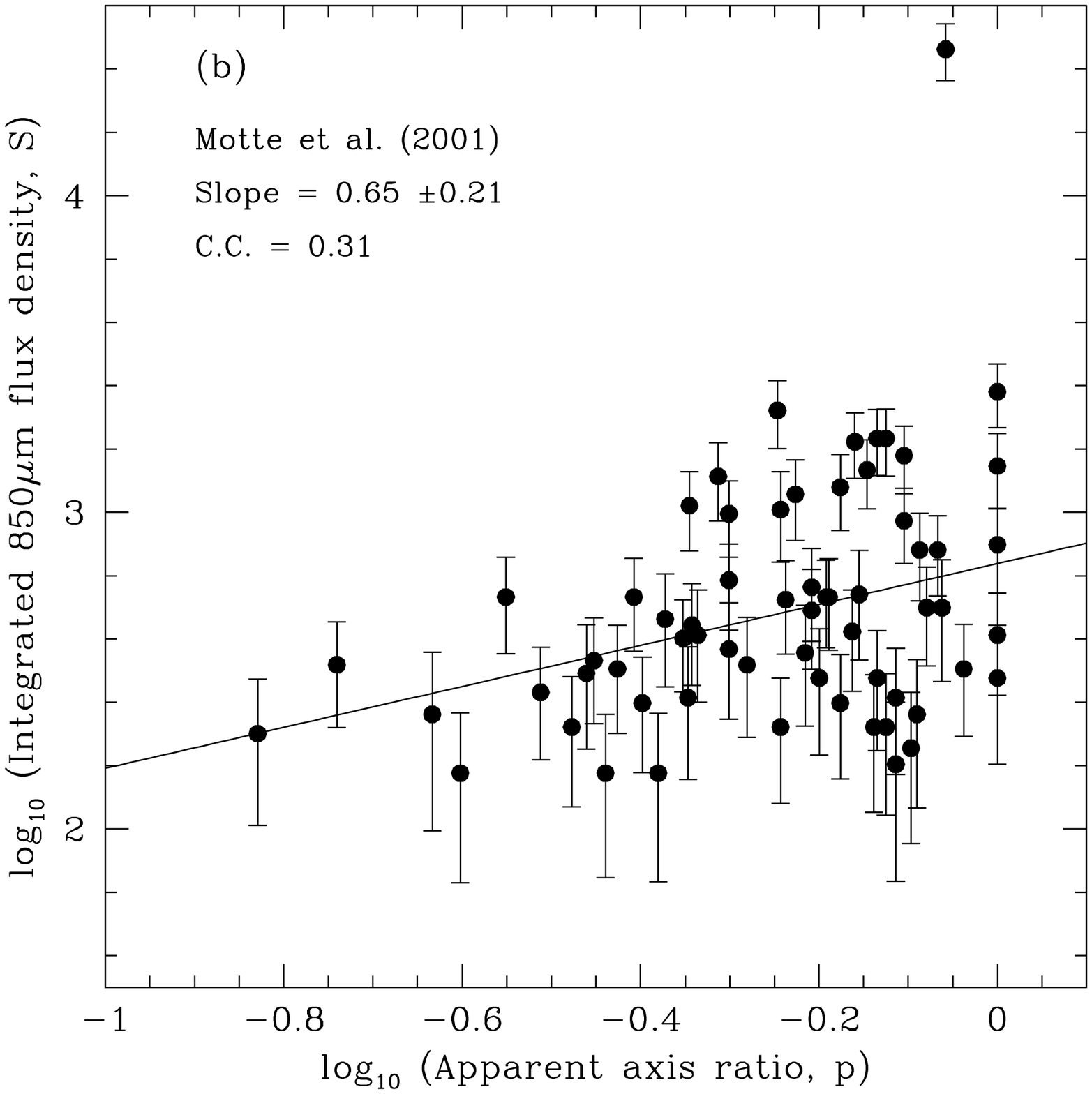}
\caption{(a) Peak intensity as a function of apparent ellipticity 
for the sample of Motte et al.\ (2001). (b) Total flux density 
as a function of apparent ellipticity for the same sample. Error bars 
in both plots reflect the calibration uncertainty of 20\% plus the rms 
noise cited in \S \ref{sec-data}. Errors in log $p$ (omitted for 
clarity) are $\pm 0.05$ in both plots.}
\efig

\bfig
\epsscale{0.55}
\plotone{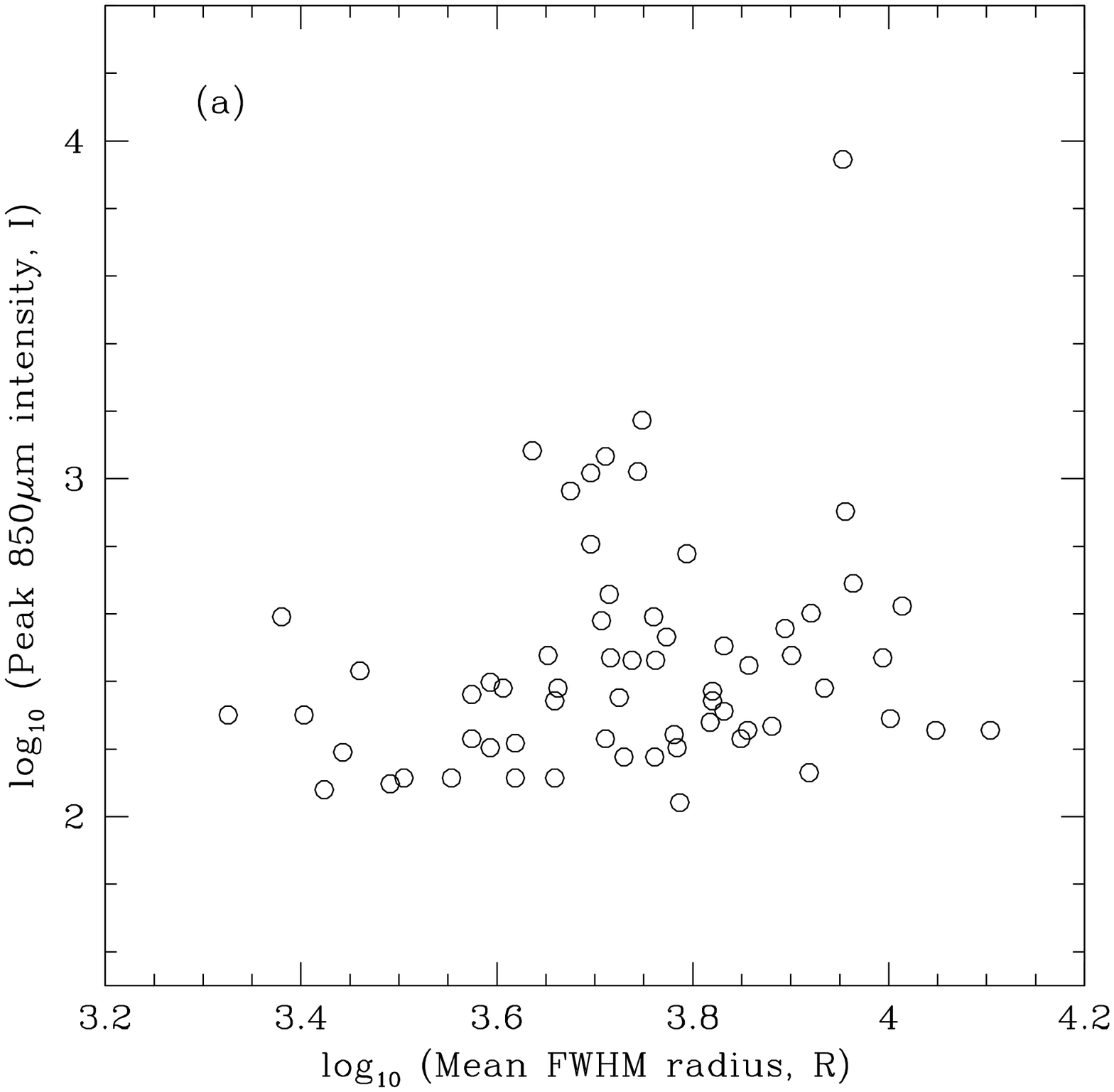}
\vspace{1cm}
\epsscale{0.55}
\plotone{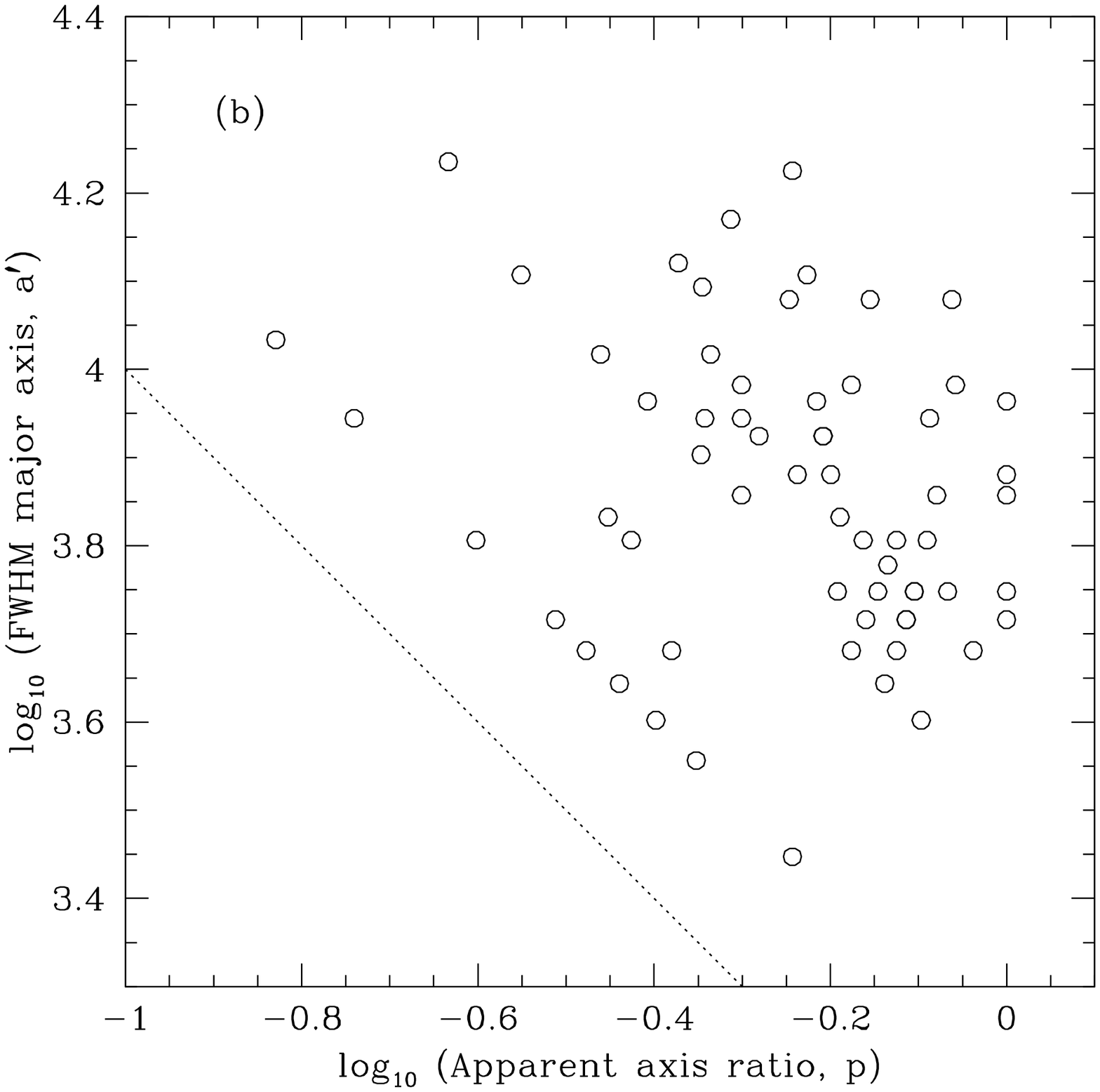}
\caption{(a) Peak intensity as a function of FWHM radius for the 
sample of M01. (b) FWHM major axis as a function of apparent 
ellipticity for the same sample. The dotted line has a slope of $-1$.}
\efig

\bfig
\epsscale{0.55}
\plotone{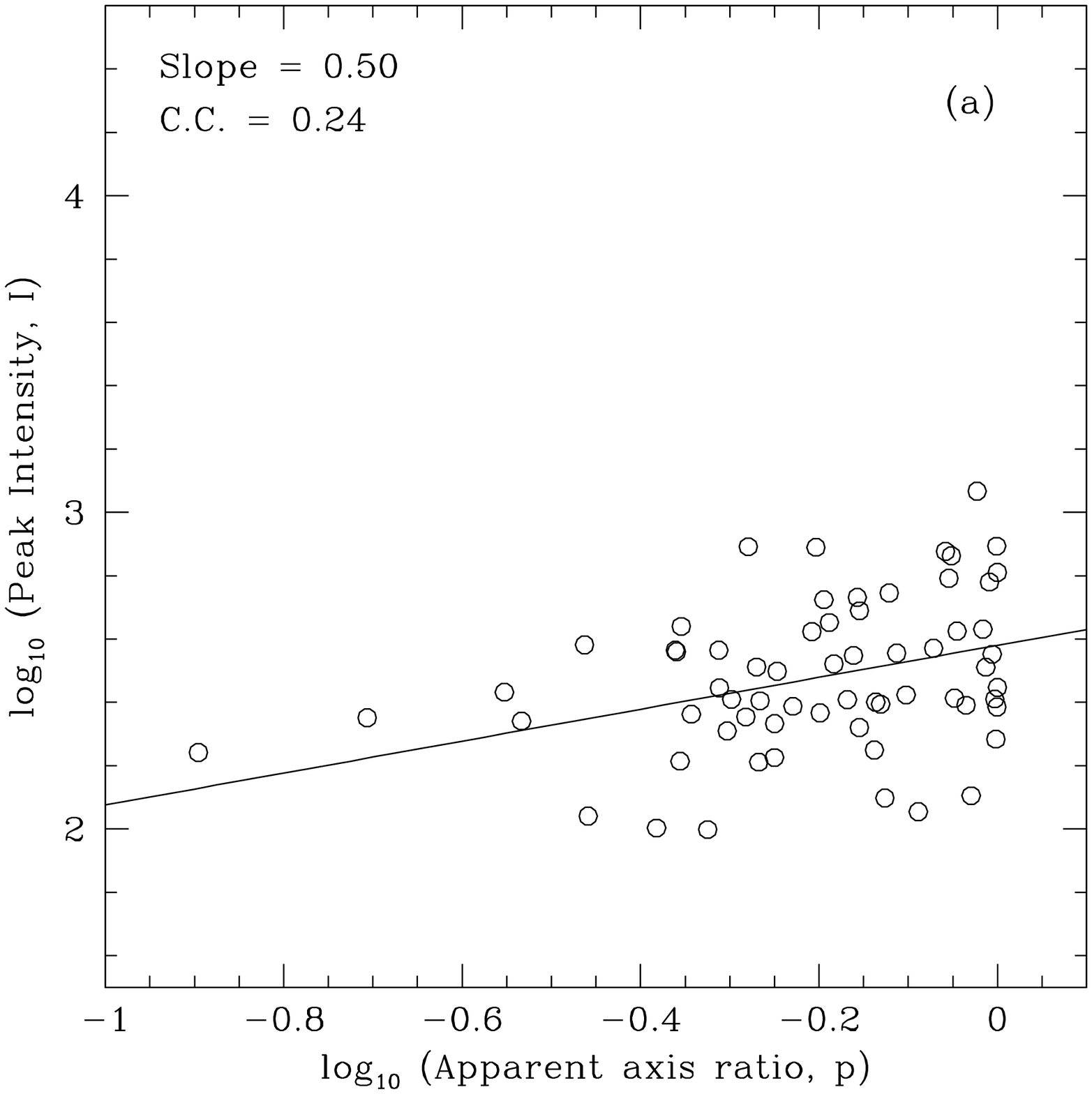}
\vspace{1cm}
\epsscale{0.55}
\plotone{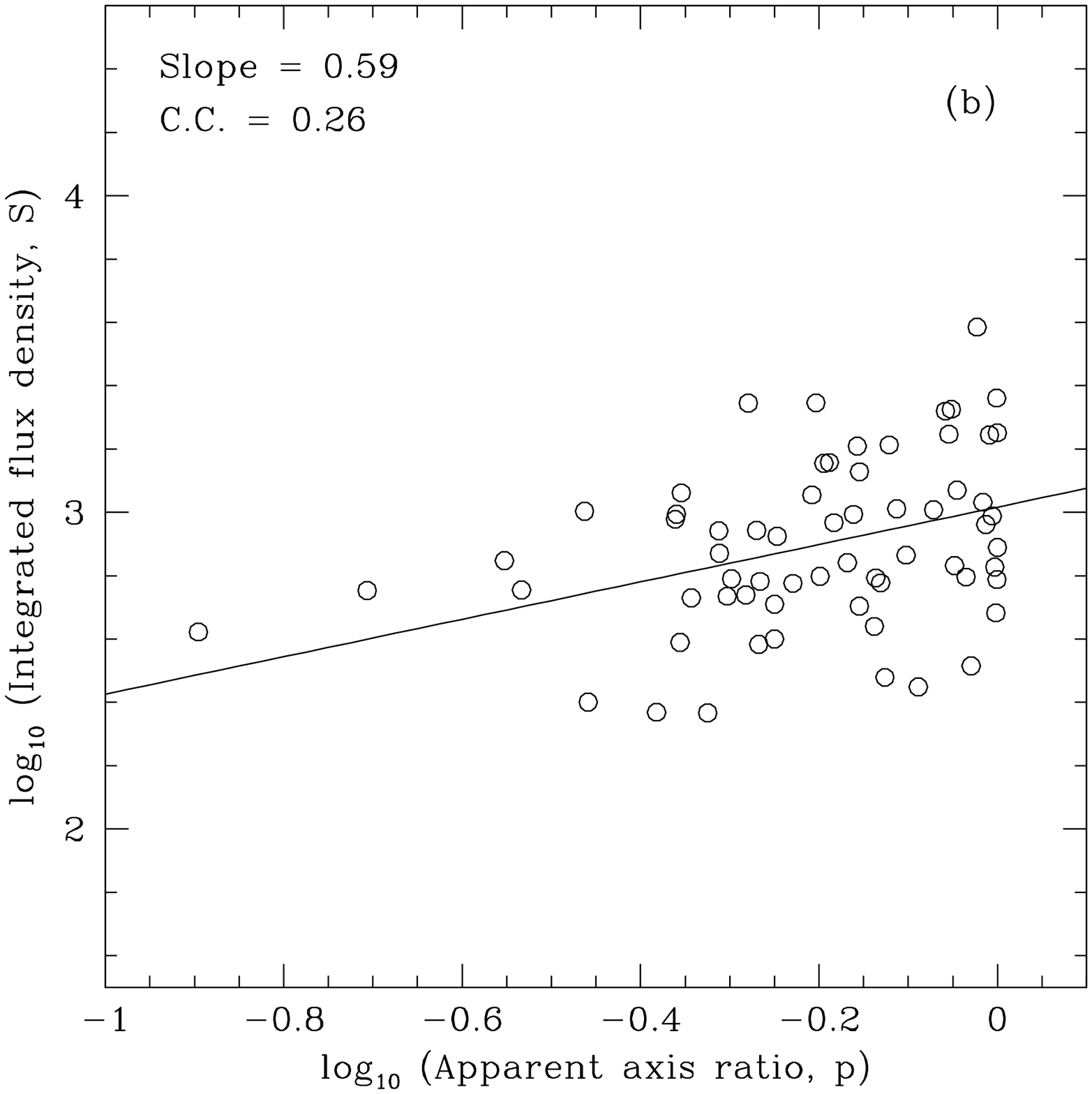}
\caption{(a) One realization from the suite of simulations of $I$ vs.\ 
$p$ described in \S \ref{sec-simul}. The correlation coefficient and 
slope of the least-squares best fit straight line are indicated at 
upper left. (b) The corresponding $S$ vs.\ $p$ plot for the same 
simulated dataset.}
\efig

\bfig
\epsscale{1.0}
\plotone{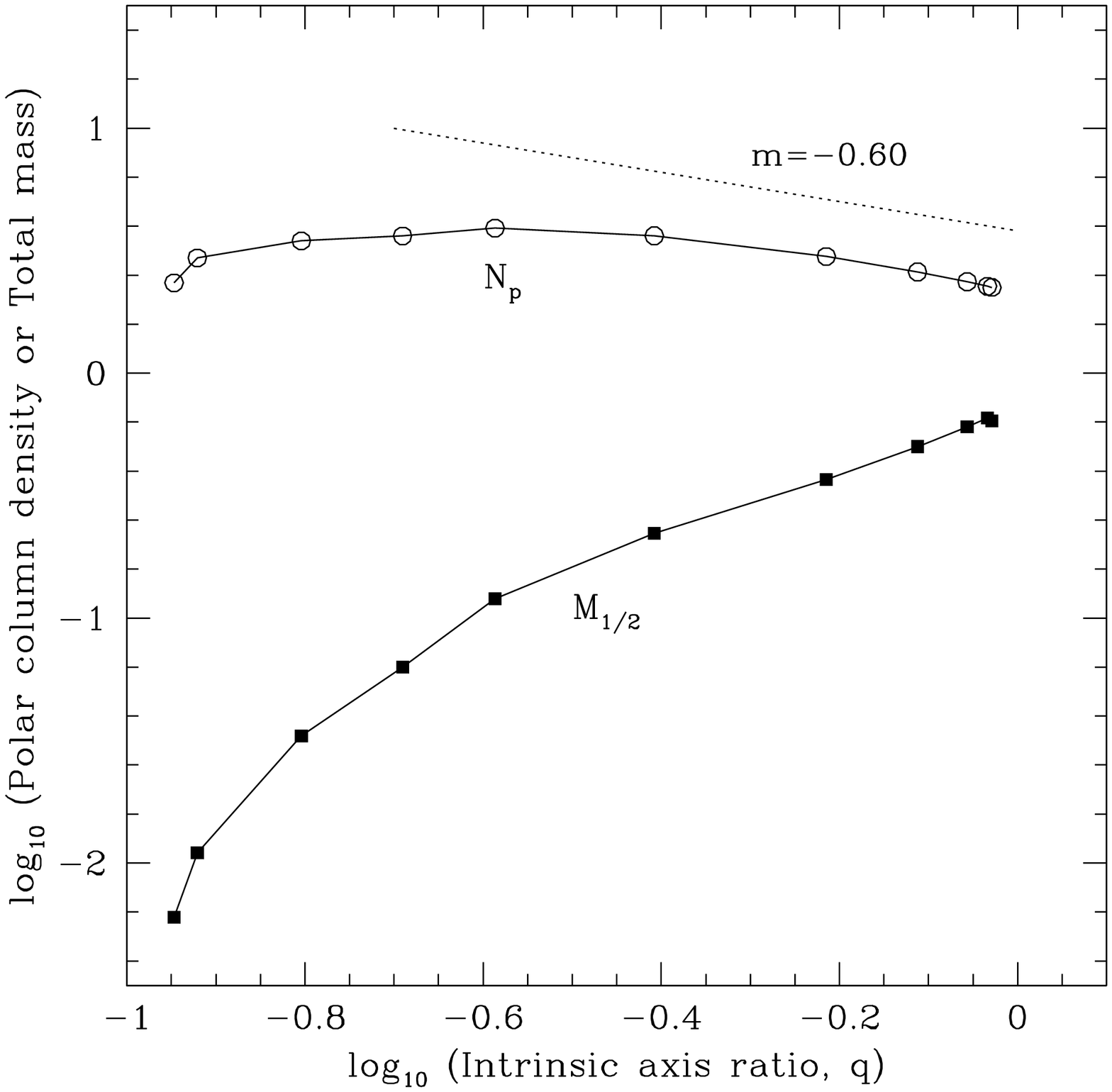}
\caption{Plot of the polar column density $N_p$ (open circles) and 
mass $M_{1/2}$ (filled squares) of cores versus intrinsic axis ratio 
in the model 
of Curry (2000). See the text of \S \ref{sec-theory} for definitions 
of these quantitities. The dotted line corresponds to the best-fit 
prolate simulations of log $I$ vs.\ log $p$, and has a slope of $-0.60$.} 
\efig

\clearpage
\begin{deluxetable}{rrrrrrrrrrr} 
\tablecolumns{11} 
\tablewidth{0pc} 
\tablecaption{Simulated log $I$ versus log $p$} 
\tablehead{ 
\multicolumn{4}{c}{}   &  \multicolumn{3}{c}{Oblate} &   
\colhead{}   & \multicolumn{3}{c}{Prolate} \\ 
\cline{5-7} \cline{9-11} \\ 
\colhead{$m$} &  \multicolumn{3}{c}{} & \colhead{Slope}   & \colhead{}    
& \colhead{Rank C.C.}    & \colhead{}
& \colhead{Slope}   & \colhead{} & \colhead{Rank C.C.}}
\startdata 
\sidehead{$\sigma$ (log $I_0$) = 0}
$-2$ & & & & $-1.68\pm 0.44$ & & $-0.30\pm 0.08$ & & \nodata & & \nodata \\ 
$-1.5$ & & & & $-1.50\pm 0.34$ & & $-0.33\pm 0.08$ & & \nodata & & \nodata \\ 
$-1$ & & & & $-1.32\pm 0.25$ & & $-0.37\pm 0.08$ & & $0.21\pm 0.15$ &
& $0.16\pm 0.08$ \\ 
$-0.5$ & & & & $-1.15\pm 0.16$ & & $-0.42\pm 0.08$ & & $0.59\pm 0.12$ & & 
$0.36\pm 0.07$ \\ 
0 & & & & $-0.97\pm 0.12$ & & $-0.46\pm 0.07$ & & $0.97\pm 0.11$ & &
$0.48\pm 0.07$ \\ 
0.5 & & & & $-0.79\pm 0.15$ & & $-0.39\pm 0.07$ & & $1.34\pm 0.13$ & &
$0.53\pm 0.06$ \\ 
1 & & & & $-0.62\pm 0.23$ & & $-0.28\pm 0.08$ & & $1.72\pm 0.17$ & &
$0.54\pm 0.07$ \\ 
1.5 & & & & \nodata & & \nodata & & $2.10\pm 0.22$ & & $0.54\pm 0.07$ \\ 
2 & & & & \nodata & & \nodata & & $2.48\pm 0.27$ & & $0.53\pm 0.07$ \\ 
\sidehead{$\sigma$ (log $I_0$) = 0.15}
$-2$ & & & & $-1.66\pm 0.46$ & & $-0.28\pm 0.08$ & & \nodata & & \nodata \\ 
$-1.5$ & & & & $-1.48\pm 0.36$ & & $-0.31\pm 0.08$ & & \nodata & & \nodata \\ 
$-1$ & & & & $-1.31\pm 0.27$ & & $-0.34\pm 0.08$ & & $0.20\pm 0.17$ &
& $0.12\pm 0.08$ \\ 
$-0.5$ & & & & $-1.14\pm 0.19$ & & $-0.37\pm 0.07$ & & $0.58\pm 0.14$ & & 
$0.29\pm 0.08$ \\ 
0 & & & & $-0.97\pm 0.16$ & & $-0.39\pm 0.07$ & & $0.96\pm 0.13$ & &
$0.41\pm 0.07$ \\ 
0.5 & & & & $-0.79\pm 0.18$ & & $-0.33\pm 0.07$ & & $1.34\pm 0.15$ & &
$0.47\pm 0.07$ \\ 
1 & & & & $-0.62\pm 0.25$ & & $-0.25\pm 0.08$ & & $1.73\pm 0.18$ & &
$0.50\pm 0.07$ \\ 
1.5 & & & & \nodata & & \nodata & & $2.11\pm 0.22$ & & $0.51\pm 0.07$ \\ 
2 & & & & \nodata & & \nodata & & $2.49\pm 0.26$ & & $0.51\pm 0.07$ \\ 
\enddata 
\end{deluxetable} 

\begin{deluxetable}{rrrrrrrrrrr} 
\tablecolumns{11} 
\tablewidth{0pc} 
\tablecaption{Simulated log $S$ versus log $p$} 
\tablehead{ 
\multicolumn{4}{c}{}   &  \multicolumn{3}{c}{Oblate} &   
\colhead{}   & \multicolumn{3}{c}{Prolate} \\ 
\cline{5-7} \cline{9-11} \\ 
\colhead{$m$} &  \multicolumn{3}{c}{} & \colhead{Slope}   & \colhead{}    
& \colhead{Rank C.C.}    & \colhead{}
& \colhead{Slope}   & \colhead{} & \colhead{Rank C.C.}}
\startdata 
\sidehead{$\sigma$ (log $I_0$) = 0}
$-2$ & & & & $-1.96\pm 0.51$ & & $-0.30\pm 0.08$ & & \nodata & & \nodata \\ 
$-1.5$ & & & & $-1.75\pm 0.39$ & & $-0.34\pm 0.08$ & & \nodata & & \nodata \\ 
$-1$ & & & & $-1.55\pm 0.28$ & & $-0.39\pm 0.08$ & & $0.25\pm 0.16$ &
& $0.17\pm 0.08$ \\ 
$-0.5$ & & & & $-1.34\pm 0.18$ & & $-0.45\pm 0.07$ & & $0.69\pm 0.12$ & & 
$0.40\pm 0.07$ \\ 
0 & & & & $-1.13\pm 0.12$ & & $-0.50\pm 0.07$ & & $1.13\pm 0.11$ & &
$0.52\pm 0.06$ \\ 
0.5 & & & & $-0.93\pm 0.17$ & & $-0.42\pm 0.07$ & & $1.57\pm 0.14$ & &
$0.56\pm 0.06$ \\ 
1 & & & & $-0.72\pm 0.26$ & & $-0.30\pm 0.08$ & & $2.01\pm 0.19$ & &
$0.56\pm 0.06$ \\ 
1.5 & & & & \nodata & & \nodata  & & $2.46\pm 0.25$ & & $0.55\pm 0.07$ \\ 
2 & & & & \nodata & & \nodata & & $2.90\pm 0.31$ & & $0.54\pm 0.07$ \\ 
\sidehead{$\sigma$ (log $I_0$) = 0.15}
$-2$ & & & & $-1.94\pm 0.53$ & & $-0.29\pm 0.08$ & & \nodata & & \nodata \\ 
$-1.5$ & & & & $-1.74\pm 0.42$ & & $-0.31\pm 0.08$ & & \nodata & & \nodata \\ 
$-1$ & & & & $-1.53\pm 0.31$ & & $-0.35\pm 0.07$ & & $0.23\pm 0.19$ &
& $0.13\pm 0.08$ \\
$-0.5$ & & & & $-1.33\pm 0.21$ & & $-0.39\pm 0.07$ & & $0.68\pm 0.15$ & & 
$0.31\pm 0.07$ \\ 
0 & & & & $-1.13\pm 0.17$ & & $-0.41\pm 0.07$ & & $1.12\pm 0.14$ & &
$0.43\pm 0.07$ \\ 
0.5 & & & & $-0.93\pm 0.20$ & & $-0.35\pm 0.07$ & & $1.57\pm 0.16$ & & $0.49\pm0.07$ \\ 
1 & & & & $-0.72\pm 0.28$ & & $-0.26\pm 0.07$ & & $2.02\pm 0.20$ & &
$0.52\pm 0.07$ \\ 
1.5 & & & & \nodata & & \nodata & & $2.47\pm 0.25$ & & $0.52\pm 0.07$ \\ 
2 & & & & \nodata & & \nodata & & $2.91\pm 0.30$ & & $0.52\pm 0.07$ \\ 
\enddata 
\end{deluxetable} 

\end{document}